\documentclass[preprintnumbers,amsmath,nofootinbib]{article}
\usepackage{bm}
\usepackage{graphicx,epsf}
\usepackage{pstricks}
\usepackage{amssymb}

\newcommand\ep{\eta_{\phi}}

\newcommand\es{\eta_{\sigma}}

\newcommand\ee{\end{equation}}

\newcommand\be{\begin{equation}}

\newcommand\eea{\end{eqnarray}}

\newcommand\bea{\begin{eqnarray}}

\newcommand\sub[1]{_{\rm #1}}

\newcommand\SUB[1]{_{\rm #1}}

\newcommand\fnl{f\sub{NL}}

\newcommand\eq[1]{Eq.~(\ref{#1})}

\begin{document}
    \title{Non-gaussianity for a Two Component Hybrid Model of Inflation}
    \author{Laila Alabidi\footnote{electronic address: l.alabidi@lancaster.ac.uk}\\ Physics Department, Lancaster University,LA1 4YB}
    \maketitle
    \begin{abstract}
    We consider a two component hybrid inflation model, in which two fields drive
    inflation. Our results show that this model generates an
    observable non-gaussian contribution to the curvature
    spectrum, within the limits allowed by the recent WMAP year 3
    data. We show that if one field has a mass $\ep<0$, and an initial value $\phi<0.06M\sub{pl}$
    while the other field has a mass $\es>0$, and initial field value $0.5M\sub{pl}<\sigma\leq{}M\sub{pl}$ then
    the non-gaussianity is observable with $1\lesssim\fnl<1.5$,
    but that $\fnl$ becomes much less than
    the observable limit should we take both masses to have the same sign, or if we loosened the
    constraints on the initial field values.
    \end{abstract}

    \section{Introduction}

        \par{}We consider hybrid inflation driven by two scalar
        fields $\phi$ and $\sigma$, with the potential:

        \be\label{pot}
            V=V_0\left(1+\frac{\ep}{2}\phi^2+\frac{\es}{2}\sigma^2\right)
        \ee

        where as usual the vacuum expectation value of the potential $V_0$ dominates, $\ep, \es$ are the masses-squared of the
        $\phi$ and $\sigma$ fields respectively and may have either sign. We take $M\sub{pl}=1$ throughout.
        This is similar to the case considered in Ref.\cite{LR,ZRL}, with the
    $\sigma=0$ trajectory, and since the corresponding
        $|f\SUB{NL}|$ was found to be much less than $1$, we aim to derive the non-gaussianity generated by a more general trajectory. We calculate the
        curvature perturbation and the resulting non-gaussianity using the
        $\Delta{}N$ formalism \cite{NG} (which is equivalent to using second-order cosmological perturbation theory as described in \cite{MW}), for which:

        \bea\label{curv}
            \zeta(t,\textbf{x})&=&\delta{}N\nonumber\\
                                &=&N_i\delta\phi_i+\frac{1}{2}N_{ij}\delta\phi_i\delta\phi_j
        \eea

        Here $N$ is the unperturbed number of $e-$folds from an initial
        time with field values $\phi_i$
         to a final time with energy density $\rho$, $N_i=\frac{\partial{}N}{\partial\phi_i}$ and $N_{ij}=\frac{\partial^2N}{\partial\phi_id\phi_j}$
         . We take the final time to be just before the end of inflation \footnote{by specifying a mechanism to end inflation, we could also
         calculate the contribution from ending as described in \cite{endinf1}, but
         we do not consider this case here.},
         and assume slow roll so that $\rho=V$.

        \par{}We measure the non-gaussianity via the amplitude
        of the bi-spectrum:

        \be\label{ng}
            \frac{3}{5}f\SUB{NL}=\frac{1}{2}\frac{N_iN_{ij}N_j}{\left(N_nN_n\right)^2}+4\ln(kaH)\mathcal{P}_{\zeta}\frac{N_{ij}N_{jk}N_{ki}}{\left(N_mN_m\right)^3}
        \ee

        where $\ln(kaH)\sim1$, $k^{-1}$ is the scale under consideration, $a$ is the scale factor, $H=\dot{a}/a$ is the Hubble
        parameter,
        and  $\mathcal{P}_{\zeta}$ is the spectrum of curvature
        perturbations. This expression assumes that the
        $\delta\phi_i$ are gaussian and applies even if
        $|f\SUB{NL}|\gtrsim{}1$ \cite{LZ,VW}, the condition
        for $f\SUB{NL}$ to eventually be observed.

    \section{The Derivatives}\label{der}

    The amplitude of curvature perturbations defined in \eq{curv} is related to the amount of
    expansion that occurs between an initial time, corresponding to a
    flat slice of space-time
    and a final time, corresponding to a space time slice of uniform energy density. Following the procedure of
    \cite{LR} (which is equivalent to that used in Ref.\cite{VW}), the respective variations of the
    fields $\phi$ and $\sigma$ are defined purely by their
    initial values (denoted here as simply $\phi$ and $\sigma$) and the amount of
    inflation.

    We use the slow roll equation:

    \be\label{SR1}
        3H\dot{\phi_n}+\frac{\partial{}V_n(\phi_n)}{\partial\phi_n}=0
    \ee

    where $\dot{\phi_n}=d\phi_n/dt$. Rearranging this equation, and
    integrating over the period of inflation we get:

    \bea
        \int_{\phi_{n*}}^{\phi_n}\frac{d\phi_n}{\phi_n}&=&-\int_{t^*}^{t}\frac{\ep{}V_0}{3H}dt\\
        &=&-\int_{0}^{N}\frac{\ep{}V_0}{3H^2}dN
    \eea
    where we have used $dN/dt\simeq{}H$ (valid for slow roll), and the $*$ denotes the time when cosmological scales left the horizon.

    \par{}Recalling that $3H^2=\sum_nV(\phi_n)\simeq{}V_0$ we have
    $\phi(N)=\phi\exp(-\ep{}N)$ and $\sigma(N)=\sigma\exp(-\es{}N)$.
    Substituting these equations into \eq{pot} we have:

   \be\label{pot2}
      V=V_0\left(1-\frac{\ep}{2}\phi^2\exp(-2\ep{}N)-\frac{\es}{2}\sigma^2\exp(-2\es{}N)\right)
    \ee

    differentiating \eq{pot2} with respect to $\phi$, while
    recalling that $V_i=0$ and rearranging:

    \be\label{np}
        \frac{\partial{N}}{\partial\phi}=N_{\phi}=\frac{
        \ep\phi\exp(-2N\ep)}{\beta}
    \ee

    where

    \be\label{beta}
    \beta(\phi,\sigma,N)=\ep^2\phi^2\exp(-2N\ep)+\es^2\sigma^2\exp(-2N\es)
    \ee

    By differentiating \eq{pot2} with respect to  $\sigma$ we get:

    \be\label{ns}
        N_{\sigma}=\frac{\es\sigma\exp(-2N\es)}{\beta}
    \ee

    Defining:

    \be\label{gamma}
    \gamma(\phi,\sigma,N)=\ep^3\phi^2\exp(-2N\ep)+\es^3\sigma^2\exp(-2N\es)
    \ee

    we get:

    \be\label{npp}
        N_{\phi\phi}=\frac{\ep\exp(-2N\ep)}{\beta}-4\frac{\ep^3\phi^2\exp(-4N\ep)}{\beta^2}+\frac{2\gamma}{\beta^3}\ep^2\phi^2\exp(-4N\ep)
    \ee

    \be\label{nss}
        N_{\sigma\sigma}=\frac{\es\exp(-2N\es)}{\beta}-4\frac{\es^3\sigma^2\exp(-4N\es)}{\beta^2}+\frac{2\gamma}{\beta^3}\es^2\sigma^2\exp(-4N\es)
    \ee

    \par{}and

    \be\label{nps}
        N_{\phi\sigma}=N_{\sigma\phi}=\frac{2\ep\phi\es\sigma\exp(-2N(\ep+\es))}{\beta^2}\left(\frac{\gamma}{\beta}-(\ep+\es)\right)
    \ee

    \section{Curvature Perturbation}

        Substituting the equations from Section \ref{der} into
        \eq{curv} we find that the curvature perturbation is given by the equation below. At first order, $\zeta$ is separable in terms of
        the individual field perturbations $\delta\phi$ and $\delta\sigma$, and
        involves a cross term $(\delta\phi\delta\sigma)$ at second order as well as the pure
        $(\delta\phi_i)^2$.

        \bea\label{zeta}
        \zeta&=&\frac{1}{2\beta}\Bigg\{2\ep\phi\exp(-2N\ep)\delta\phi+2\es\sigma\exp(-2N\es)\delta\sigma\nonumber\\
        &+&\left[\ep\exp(-2N\ep)-4\frac{\ep^3\phi^3}{\beta}\exp(-4N\ep)+2\frac{\gamma}{\beta^2}\ep^2\phi^2\exp(-4N\ep)\right](\delta\phi)^2\nonumber\\
        &+&\left[\es\exp(2N\es)-4\frac{\es^3\sigma^3}{\beta}\exp(-4N\es)+2\frac{\gamma}{\beta^2}\es^2\sigma^2\exp(-4N\es)\right](\delta\sigma)^2\nonumber\\
        &+&2\left[2\frac{\ep\phi\es\sigma\exp(-2N(\es+\ep))}{\beta}\left(\frac{\gamma}{\beta}-(\ep+\es)\right)\right]\delta\phi\delta\sigma\Bigg\}\nonumber\\
        \eea

    \section{ $f\SUB{NL}$}

    Substituting Eqs.(\ref{np}) to (\ref{nps}) into
    \eq{ng} we find that the non-gaussian contribution to the curvature spectrum is given by:%\newpage

       \bea\label{ng2}
        \frac{3}{5}f\SUB{NL}&=&\frac{1}{2\left[\ep^2\phi^2\exp(-4N\ep)+\es^2\sigma^2\exp(-4N\es)\right]^2}\nonumber\\
        &&\times\Bigg\{-\beta\left[-\es^3\sigma^2\exp(-6N\es)-\ep^3\phi^2\exp(-6N\ep)\right]\nonumber\\
        &&+2\left[-2\es^5\sigma^4\exp(-8N\es)-2\ep^5\phi^4\exp(-8N\ep)\right.\nonumber\\
        &&\left.-2(\ep+\es)\ep^2\es^2\phi^2\sigma^2\exp(-4N(\es+\ep))\right]\nonumber\\
        &&+2\frac{\gamma}{\beta}\left[\es^4\sigma^4\exp(-8N\es)+\ep^4\phi^4\exp(-8N\ep)\right.\nonumber\\
        &&\left.+2\ep^2\es^2\phi^2\sigma^2\exp(-4N(\es+\ep))\right]\Bigg\}+\cdots
    \eea

    \par{}We found that the second term is negligible with respect to the first, so we do not include it
    here.

    \section{The Positive-Negative Mass Combination}

    In this section we focus on taking $\es>0$ while $\ep<0$.
    In this case the $\sigma$ field is pushing the inflaton towards
    the origin while the $\phi$
    field is pulling it away.

    \par{}To first order in field perturbations, we find that the
    curvature perturbation \eq{curv} is dominated by the
    fluctuations in the \emph{negative} mass field, since the
    fluctuations $\delta\sigma$ are exponentially damped.

    \be
    \zeta=\frac{1}{\ep\phi}\delta\phi-\frac{2\es\sigma}{\ep^2\phi^2}\exp(2N(\ep-\es))\delta\sigma
    \ee

    \par{}This argument \emph{could} be used to simplify the
    non-gaussian term (\ref{ng2}) with only a small loss of
    precision, especially for the case where $\es\sim{}1$, and would
    reproduce the $\fnl$ for a single field model. However
    this is not the case for $\es\ll{}1$, as then the
    exponent $2N\es\sim{}1$, and it becomes interesting to consider
    the affect of the $\delta\sigma$ fluctuations on observation.

    \section{The Spectral Index}

    For a multi-field inflaton the dominant term in the spectral index is defined
    by \cite{LL}:

    \[
    n-1=2\frac{N_aN_bV_{ab}}{VN_dN_d}\nonumber\\
    \]

    so by substituting Eqs. (\ref{np}) and (\ref{ns}), and using
    \eq{pot} to calculate $V_{ab}$ we get:

    \be\label{spec}
    n-1=2\frac{\ep^3\phi^2\exp(-4N\ep)+\es^3\sigma^2\exp(-4N\es)}{\ep^2\phi^2\exp(-4N\ep)+\es^2\sigma^2\exp(-4N\es)}
    \ee

    \par{}From the recent WMAP year three data \cite{wmap3}
    the range of allowed $n-1$ for a negligible tensor fraction $r$ at the $2\sigma$ limit is:

    \be\label{limit}
    -0.083<n-1<-0.02
    \ee

    \par{}If we want the spectral index to fall within observational
    limits regardless of initial conditions,
    then we require the spectral index given in \eq{spec} to fall within the observational bounds \eq{limit} independently of the initial field values.
    For the case where both $\es,\ep<0$ then since \eq{spec} is a
    weighted sum of $2\eta_i$ we require:

    \be\label{limit2}
    |\eta_i|<0.041
    \ee

    \par{}However, for the case where $\ep<0$ and $\es>0$, then as long as $\ep$ satisfies
    the limits set by \eq{limit2} then $\es$ can take on any value less than $1$, since for
    large values of $\es$ terms within the equations including it are exponentially
    damped. Also note that for positive field masses, $n-1>0$ which
   is ruled out by observation.

    \section{Results for the Non-Gaussianity}

    \par
\begin{figure}
            \centering\includegraphics[angle=270,width=4in]{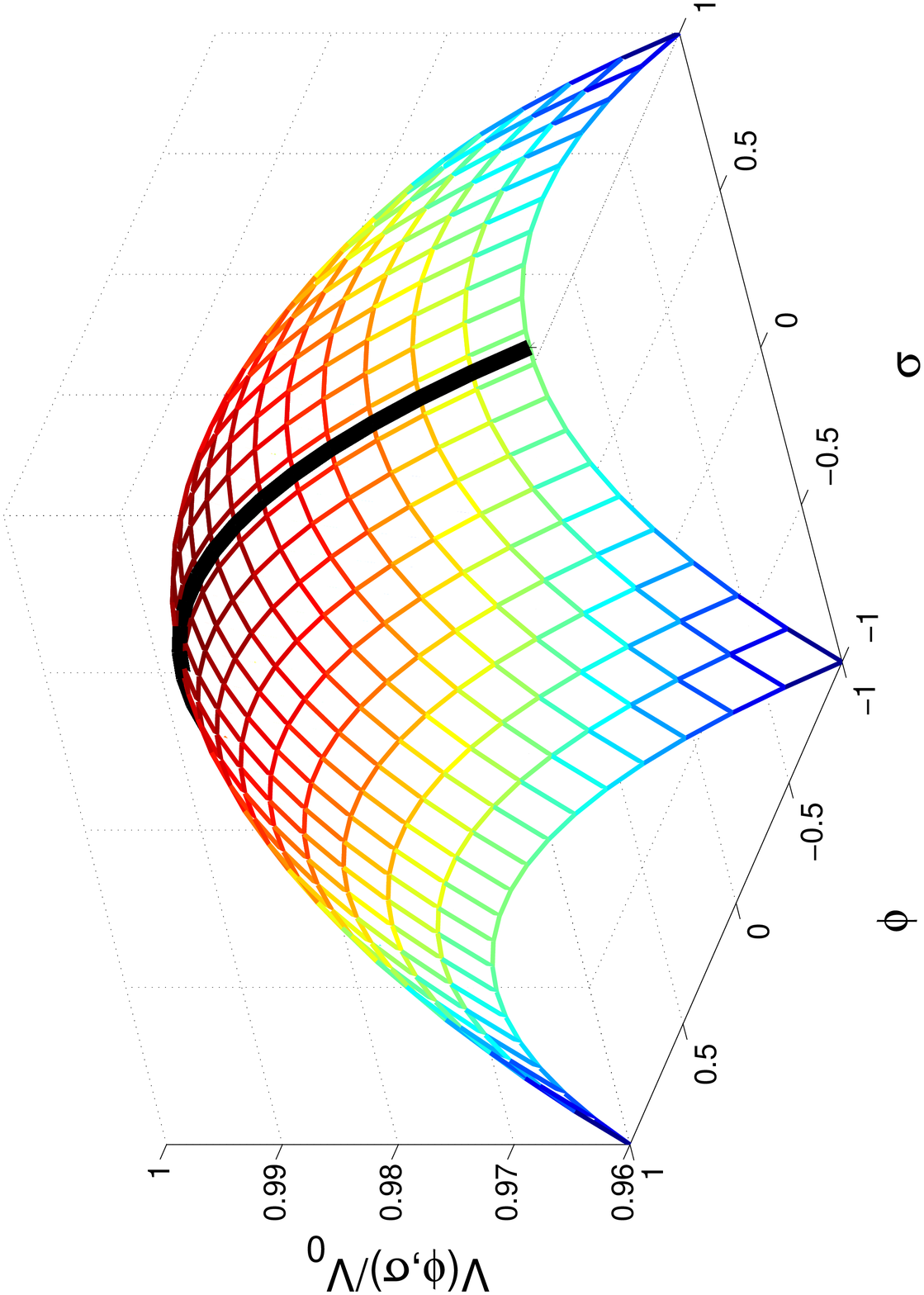}
            \caption{The potential \eq{pot} for $\ep,\es>0$, in which the inflaton starts at a maximum and is pushed \emph{away} from the origin by \emph{both}
            fields. The solid black line corresponds to the unperturbed case $\sigma=0$}
            \label{neg-neg}
    \end{figure}

{}We begin by analyzing \eq{ng2} for
    $N=50$. Keeping within the range of $\eta_i$ allowed by
    \eq{limit}, we have analysed the
    potential for initial field values ranging between
    $(0 - 1)$, and values of the masses ranging between $(-0.04 - 0)$
    for $\ep$ and $(-0.04 - 0.15)$ for $\es$. We then extracted
    the maximum value of $f\SUB{NL}$ for each combination of masses and plot the
    results in
    Fig\ref{max}. We repeat this procedure and extract $|\frac{3}{5}f\SUB{NL}|_{max}$
    for each combination of initial field values and plot these results in Fig\ref{contour_ini}. We have
    also provided illustrations of the potential for the two
    scenarios considered in this paper in
    Figs\ref{neg-neg} and \ref{pos-neg}. For the case where
    $\es\to0$ and $-0.04<\ep<0$, \eq{ng2} reproduces
    the results for a single field scenario.

    \par{}It is a far from easy task to extract any meaningful
    information from a four parameter model such as this one.
    However, both Figs. \ref{max} and \ref{contour_ini} are results of a full parameter space
    calculation of $\fnl$, but Fig.\ref{max} shows the mass dependence of $|\fnl|$, while Fig.\ref{contour_ini} shows the initial field value dependance
    of $|\fnl|$ and by analyzing the figures we can see that for large $\sigma$,
    small $\phi$, and $\es>0,\ep<0$ we get $1\lesssim\fnl<1.5$.
    Yet, if we were to consider small values of the fields
    $\phi,\sigma<0.1$, we see that $\fnl\ll1$ regardless of
    the values of the masses.

\begin{center}
    \begin{figure}[h]
        %\begin{minipage}[t]{\textwidth}
        \centering\includegraphics[angle=270,width=\linewidth,totalheight=2.5in]{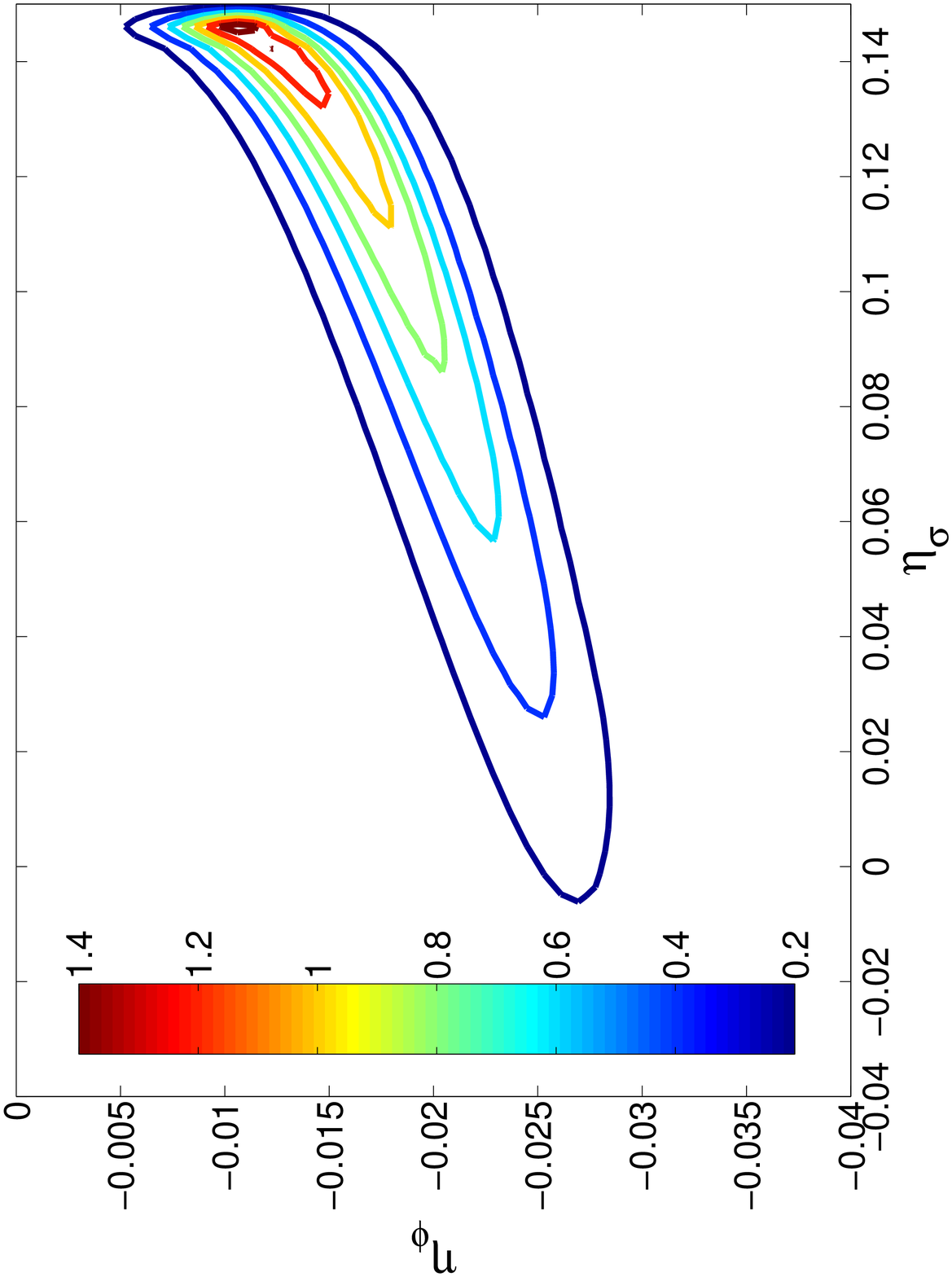}
        \caption{Contour plot of the absolute value of the maximum
        non-gaussianity generated for various combinations of masses,
        with initial conditions ranging between zero and the Planck
        mass. $|\fnl|\gtrsim1$ corresponds to the four inner
        contours and the region outside the outermost contour
        corresponds to $|\frac{3}{5}\fnl|_{max}<0.2$.
         }
        \label{max}
        %\end{minipage}\\
    \end{figure}
    \begin{figure}
        \begin{minipage}[c]{0.25\linewidth}
            \includegraphics[angle=270,width=5in,totalheight=3in]{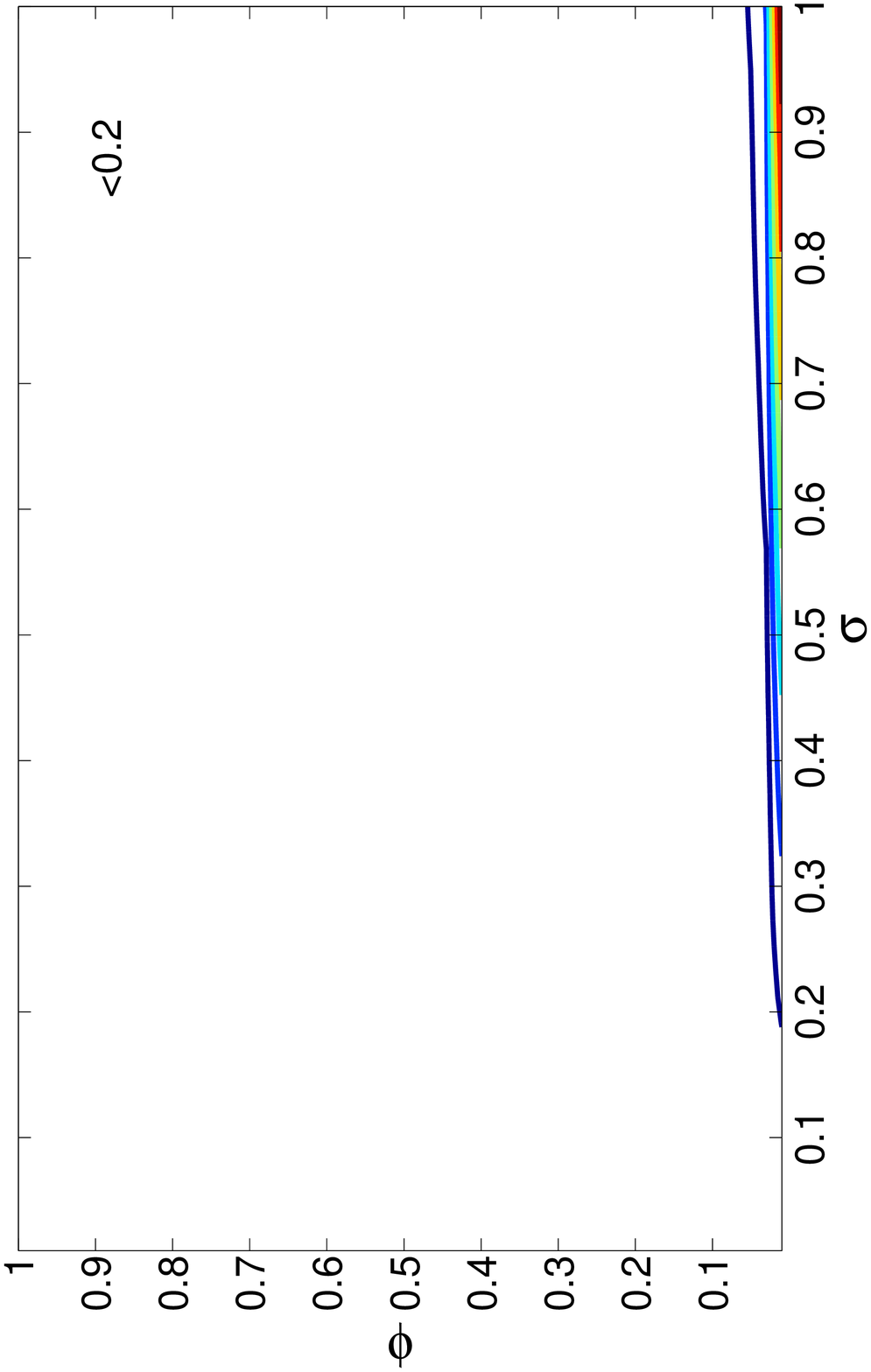}
            \end{minipage}%
        \begin{minipage}[c]{0.5\linewidth}
            \includegraphics[angle=270,width=2.5in,totalheight=2in]{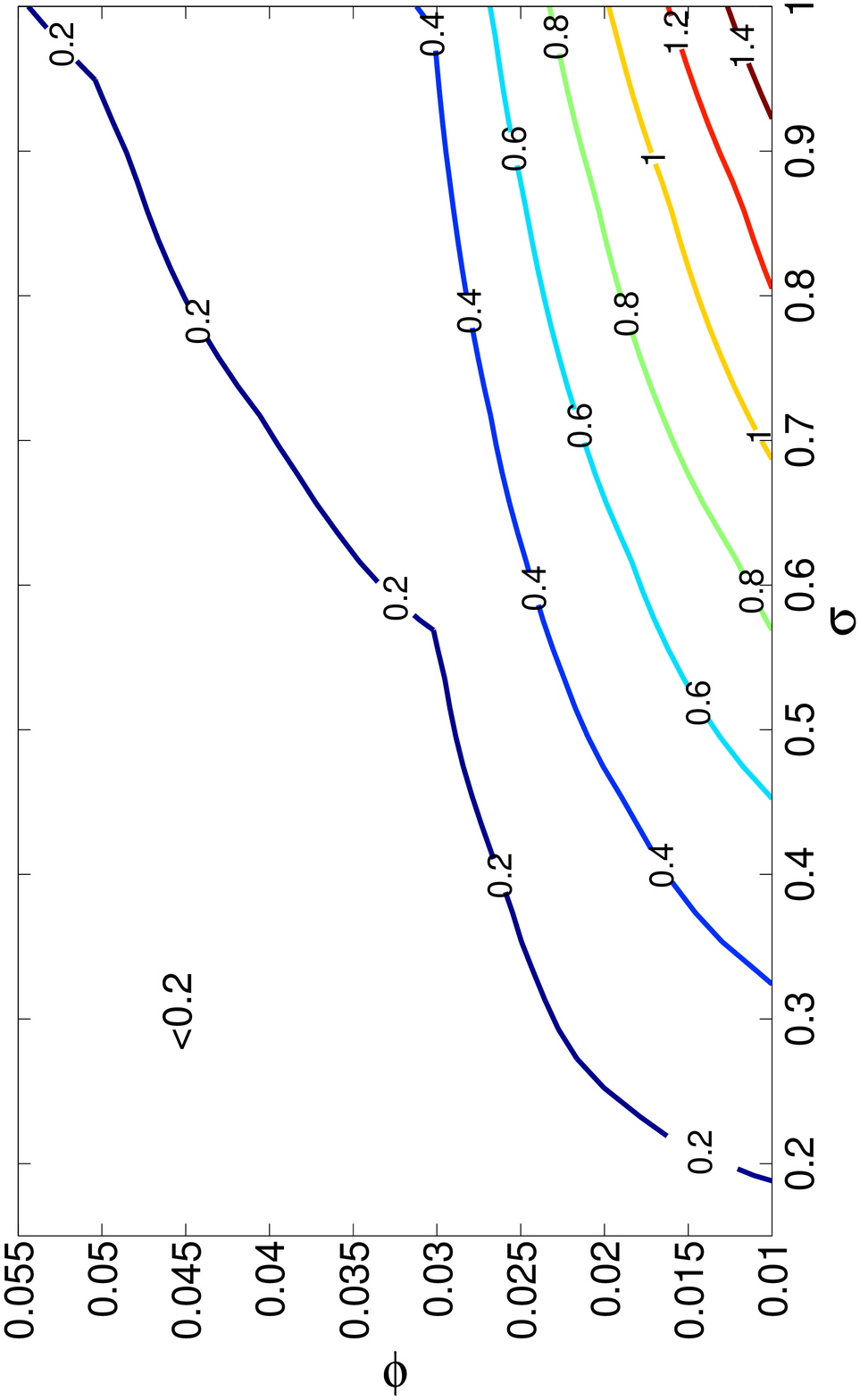}
        \end{minipage}
            \caption{Contour plot of the absolute value of the maximum non-gaussianity generated for various combinations
            of initial field values, with masses $\ep$ ranging between ($-0.04-0$) and $\es$ ranging between ($-0.04 - 0.15$).
            The embedded plot is a magnification of the region in which $|\frac{3}{5}f\SUB{NL}|>0.2$ appears, note that this
            corresponds to $\phi\ll1$ and $0.2\leq\sigma\leq1$.}
              \label{contour_ini}
    \end{figure}
\end{center}

    \begin{figure}
            \centering\includegraphics[angle=270,width=4in]{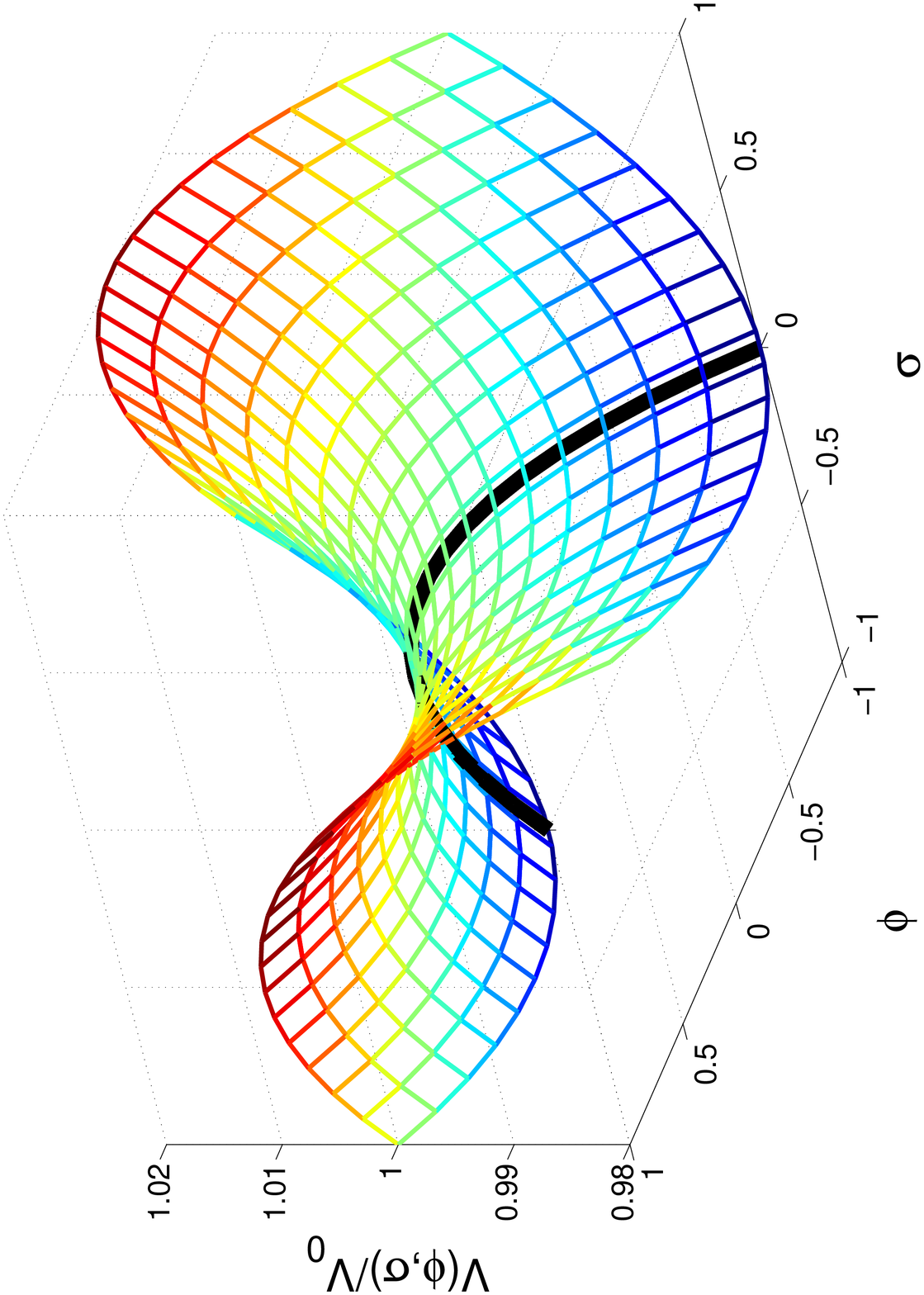}
            \caption{The potential \eq{pot} for $\ep>0$ and $\es<0$, in which the $\phi$ field pushes the inflaton
            \emph{away} from the origin while the $\sigma$ field pulls it $towards$ it. This case produces the more significant
            (yet still undetectable) values of $f\SUB{NL}$ which appear in Figs.\ref{max}\&\ref{contour_ini}. The solid black line corresponds to the unperturbed case $\sigma=0$}
            \label{pos-neg}
    \end{figure}

\newpage
   \section{Discussion}

   We summarise the main results in the table below, and by \emph{Consistent with $n-1$?}
   we mean does it satisfy the limits on the spectral index in \eq{limit}.

   \begin{center}
    \begin{tabular}{|c|c|c|c|c|c|}
    \hline
     & & & & &Consistent\\
    $\es$&$\ep$&$\sigma$&$\phi$&$|\fnl|$&with\\
    & & & & &$n-1$?\\
    \hline
    $0<\es\ll1$&$<0$&$<1$&$\ll1$&$1\lesssim|\fnl|<1.5$&yes\\
    \hline
    $<1$&$<0$&any value&any value&single field case&yes\\
    &&&&$|\fnl|\ll1$&\\
    \hline
    $<0$&$<0$&any value&any value&$|\fnl|\ll1$&yes\\
    \hline
    any value&any value&$\ll1$&$\ll1$&$|\fnl|\ll1$&yes\\
    \hline
    $>0$&$>0$&any value&any value&not&no\\
    &&&&calculated&$n-1>0$\\
    \hline
    $\to0$&$<0$&any value&any value&single field case&yes\\
    &&&&$|\fnl|\ll1$&\\
    \hline
    \end{tabular}
    \end{center}

  \par{}For the first result listed $\es\ll1$, considering that an unperturbed trajectory
  corresponds to $\sigma=0$ (solid black lines in Figs \ref{neg-neg} and \ref{pos-neg}),
  the non-gaussian contribution
  to the curvature spectrum is due to the slight deviation in the
  inflaton trajectory caused by the $\sigma$ field, and since
  $-2N\es\sim1$, the effect `survives' inflation, leaving it's
  imprint on the spectrum. In the $\es<1$ case, the roll to the
  minimum occurs rapidly at the beginning of inflation (by rapidly
  we mean on a time scale much smaller than the time scale of
  inflation), and thus the effect is negligible, and the results
  for the non-gaussianity are equivalent to a single field case
  $\frac{3}{5}\fnl\simeq\ep/2$.

  \par{}It is only fair to point out that we have not provided a
  physical motivation for this type of model with field values of order one, and since much
 effort was put into justifying multi-field chaotic models (as
summarized in \cite{inflation}), it is not clear whether we can
   write down our potential without justification.

  \section{Acknowledgements}

  I thank David H.~Lyth and Karim Malik for helpful suggestions.
  I also thank Lancaster University for the award of the studentship from the Dowager Countess
  Eleanor Peel Trust.

\end{document}